# BAlN for III-nitride UV light emitting diodes: undoped electron blocking layer

Wen Gu,[1,2,3] Yi Lu,[1,2,3] Rongyu Lin,[2] Wenzhe Guo,[2] Zi-Hui Zhang,[4] Jae-Hyun Ryou,[5] Jianchang Yan,[1,*] Junxi Wang,[1] Jinmin Li,[1] and Xiaohang Li[2,*]

[1] Research and Development Center for Solid State Lighting, Institute of Semiconductors, Chinese Academy of Sciences, Beijing 100083, China
[2] Advanced Semiconductor Laboratory, King Abdullah University of Science and Technology, Thuwal 23955-6900, Saudi Arabia
[3] Center of Materials Science and Optoelectronics Engineering, University of Chinese Academy of Sciences, Beijing 100049, China
[4] Institute of Micro-Nano Photoelectron and Electromagnetic Technology Innovation, School of Electronics and Information Engineering, Hebei University of Technology, Key Laboratory of Electronic Materials and Devices of Tianjin, 5340 Xiping Road, Beichen District, Tianjin, 300401, China
[5] Department of Mechanical Engineering, Advanced Manufacturing Institute (AMI), and Texas Center for Superconductivity at UH (TcSUH), University of Houston, Houston, TX 77204-4006, U.S.A.

*yanjc@semi.ac.cn, xiaohang.li@kaust.edu.sa

**Abstract:** The undoped BAlN electron-blocking layer (EBL) is investigated to replace the conventional AlGaN EBL in light-emitting diodes (LEDs). Numerical studies of the impact of variously doped EBLs on the output characteristics of LEDs demonstrate that the LED performance shows heavy dependence on the p-doping level in the case of the AlGaN EBL, while it shows less dependence on the p-doping level for the BAlN EBL. As a result, we propose an undoped BAlN EBL for LEDs to avoid the p-doping issues, which a major technical challenge in the AlGaN EBL. Without doping, the proposed BAlN EBL structure still possesses a superior capacity in blocking electrons and improving hole injection compared with the AlGaN EBL having high doping. This study provides a feasible route to addressing electron leakage and insufficient hole injection issues when designing UV LED structures.

## 1. Introduction

AlGaN-based (III-N) ultraviolet (UV) light-emitting diodes (LEDs) have aroused widespread interests over the past few decades due to their various potential applications in purification, bio-detection, medical treatment, next-generation data storage, and lithography [1]. As a substitute for the conventional mercury lamp, UV LEDs are potentially energy efficient, long lifetime, compact, and environmentally friendly. However, the low efficiency and output optical power of the UV LEDs have hampered their adoption in various applications [2]. The currently developed LEDs operating in UV spectral regions still suffer from relatively low external quantum efficiency (EQE) and substantial efficiency droop effect [3,4]. Reasons for the low efficiency include the insufficient hole injection into the active region [5] and severe electron leakage out of the active region originating from the large valence band offset and small conduction band offset between quantum wells (QWs) and quantum barriers (QBs) [6], respectively. However, the commonly used electron-blocking layer (EBL) possesses several concerning issues. The electronic band edge profiles can be bent because of the polarization-induced electrostatic field, which may increase the hole injection barrier and further deteriorate the output performance of the LEDs [7]. Moreover, sufficient p-doping level AlGaN EBL is preferable for blocking electrons [8]. However, the activation energy of widely used p-dopants, e.g. Mg, dramatically rises with increasing Al mole fraction in the AlGaN layer, which makes the ionization of acceptors more challenging [9]. Moreover, the diffusion of Mg atoms from the p-region to the active region is more severe in high-Al composition structures [10]. The induced Mg-related defects in multiple QWs (MQWs) will form nonradiative recombination centers, which is detrimental to the low internal quantum efficiency (IQE) [11]. Besides, the Mg-induced defect will scatter electrons, leading to a low electron mobility [12].

To address the issues associated with the electron leakage, the hole injection, and the p-EBL, various solutions in the layer structures have been proposed. A superlattice was used as the EBL to suppress electron leakage and improve the overall performance of UV LEDs [13]. A quaternary AlInGaN EBL was also employed to reduce the polarization charge density in the heterostructure interface, which facilitates the reduction in band bending of the EBL [14]. Moreover, an EBL-free UV LED structure was proposed by utilizing graded-composition AlGaN QBs to realize better electron blocking and hole injection as opposed to the conventional structure with a p-EBL [15]. Researchers also designed the hole injection layers inserted between EBL and MQWs to effectively relieve the polarization-induced valence band bending [16]. Moreover, EBLs with graded composition [17], V-shaped structures [18], two-step tapered structures [19,20] as well as polarization doped layers [21] show favorable potentials for UV LEDs. However, most of these methods could still suffer from the Mg diffusion issue.

Boron-containing III-N alloys, especially BAlN, are emerging wide-bandgap materials for optoelectronic and power devices. Recently, researchers have successfully grown BAlN/AlN and BAlN/AlGaN superlattices [22,23]. The epitaxial growth of monocrystalline wurtzite BAlN structure with boron content as high as 11% and 14.4% has been demonstrated [24]. Liu et al. have calculated the spontaneous polarization (SP) and piezoelectric (PZ) constants of BAlN using hexagonal reference structures [25]. The results also revealed that the heterointerface polarization can be modulated by adjusting the boron composition, which is beneficial for designing polarization-related electronic devices. Importantly for UV LEDs, the band alignment of BAlN/(Al)GaN heterostructure is extremely advantageous for electron confinement and hole injection. The valence and conduction band edges of $B_{0.14}Al_{0.86}N$ are reportedly 0.2 eV lower and 2.1 eV higher, respectively, than those of GaN [26,27]. Thus, the BAlN EBL is promising to supersede the conventional AlGaN EBL because of the possibility of suppressing electron leakage effectively without severely deteriorating hole injection.

In this study, motivated by the conduction and valence band offset properties, $B_{0.14}Al_{0.86}N$ is employed as an alternative to $Al_{0.3}Ga_{0.7}N$ for the EBL in UV LEDs. First, we systematically investigate the effect of p-doping in the $Al_{0.3}Ga_{0.7}N$ EBLs with various doping levels on the output performance of LEDs. The result shows that the p-doping level of an EBL has a great influence on the effective barrier heights of the conduction and valence bands. The high p-doping level in the $Al_{0.3}Ga_{0.7}N$ EBL decreases the valence band offset and increases the conduction band offset, facilitating hole injection and the confinement of electrons, respectively. Meanwhile, we further investigate the $B_{0.14}Al_{0.86}N$ EBL, which shows the same tendency on effective band barrier heights as the $Al_{0.3}Ga_{0.7}N$ EBL. However, the conduction and valence bands of $B_{0.14}Al_{0.86}N$ EBL can maintain relatively high and low offsets, respectively, even with decreasing the Mg doping concentration. Finally, we propose an innovative undoped $B_{0.14}Al_{0.86}N$ EBL for UV LED with superior performance to avert the challenging p-doping issue in high-Al composition layers.

## 2. Structures and parameters

Fig. *Fig. 1*. A schematic cross-sectional structure of UV LEDs with variously doped $Al_{0.3}Ga_{0.7}N$ or $B_{0.14}Al_{0.86}N$ EBLs. presents a schematic cross-sectional structure of AlGaN LEDs including either the conventional $Al_{0.3}Ga_{0.7}N$ EBL or the proposed $B_{0.14}Al_{0.86}N$ EBL. The conventional structures are grown on a GaN template, followed by a 3-μm-thick n-type $Al_{0.2}Ga_{0.8}N$ layer doped with silicon at a concentration of $5 \times 10^{18}$ /cm³ (n-$Al_{0.2}Ga_{0.8}N$:Si, 3 μm, [Si] = $5 \times 10^{18}$ /cm³). The active region is composed of five $Al_{0.1}Ga_{0.9}N$ (3 nm each) QWs and six $Al_{0.2}Ga_{0.8}N$ (14 nm each) QBs emitting at 340 nm. Above the last QB is a 20-nm-thick $Al_{0.3}Ga_{0.7}N$ EBL with various doping levels ($Al_{0.3}Ga_{0.7}N$:Mg, 20 nm, [Mg] = 0, $1 \times 10^{15}$, $1 \times 10^{16}$, $1 \times 10^{17}$, $1 \times 10^{18}$, $1 \times 10^{19}$, $1 \times 10^{20}$ /cm³). Then, a p-$Al_{0.2}Ga_{0.8}N$:Mg layer (100 nm, [Mg] = $2 \times 10^{19}$ /cm³) and a p-GaN layer (10 nm, [Mg] = $1 \times 10^{20}$ /cm³) are deposited in sequence. For the proposed structures, a $B_{0.14}Al_{0.86}N$ EBL with the same thickness and changes in p-doping levels as the $Al_{0.3}Ga_{0.7}N$ EBL is used. The other layers remain the same as the conventional structures. Both LED structures are designed to be $300 \times 300$ μm² in size.

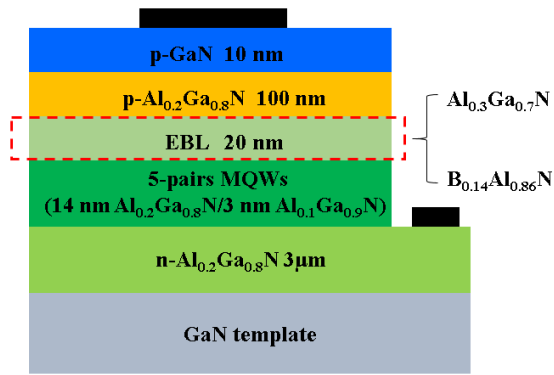

Fig. 1. A schematic cross-sectional structure of UV LEDs with variously doped $Al_{0.3}Ga_{0.7}N$ or $B_{0.14}Al_{0.86}N$ EBLs.

We assume the conduction/valence band offset ratio of $Al_{0.2}Ga_{0.8}N/Al_{0.1}Ga_{0.9}N$ MQWs is 0.7/0.3 [28]. The SP and PZ constants of $B_{0.14}Al_{0.86}N$ and $Al_xGa_{1-x}N$ (x= 0, 0.1, 0.2, 0.3) are from Refs [25,29], which have proven to be accurate in calculating the polarization of III-N materials [30]. The energy bandgap of $Al_xGa_{1-x}N$ alloys is estimated using Equation (1), where b is a bowing constant and is chosen to be 0.94 [31], x is the Al content. The bandgap of $B_{0.14}Al_{0.86}N$ is set at 5.7 eV [32].

$$E_g(Al_xGa_{1-x}N) = xE_g(AlN) + (1-x)E_g(GaN) - bx(1-x) \tag{1}$$

The Auger recombination coefficient and Shockley-Read-Hall (SRH) recombination lifetime are chosen as $1.0 \times 10^{-30}$ /cm$^3$ [33] and 50 ns, respectively. The screening factor is set to 40%, which is a commonly used value for calculating the polarization induced built-in interface charges [34]. The operating temperature and background loss are estimated to be 300 K [35] and 2000 m$^{-1}$ [36], respectively. Although the p-type $B_{0.14}Al_{0.86}N$ has not been demonstrated in the experiment yet, the acceptor activation energy of $B_{0.14}Al_{0.86}N$ should be similar to that of GaN because of their analogous valence band edge [26]. The effective mass of $B_{0.14}Al_{0.86}N$ is from ref [32]. The activation energy of GaN or $B_{0.14}Al_{0.86}N$ is supposed to be 170 meV [37], and the activation energy of $Al_xGa_{1-x}N$ is assumed to be 270 meV [38]. Generally accepted material parameters, including effective mass, electron and hole mobility values are applied to $Al_xGa_{1-x}N$ and GaN layers [39].

## 3. Effects of p-doping level in EBLs

The p-doping level of EBL is a critical factor that deserves a considerable attention in designing high-performance LED structures. To evaluate the effects of EBLs at different doping levels on the performance of LEDs, we design the EBLs with a series of Mg doping concentrations (as described in part 2). Fig. Fig. 2. Electronic band edge profiles at an injection current of 90 mA for the LED structures with (a) $Al_{0.3}Ga_{0.7}N$ EBL and (b) $B_{0.14}Al_{0.86}N$ EBL at various Mg doping concentrations. shows the electronic band edge profiles for the LED structures with an $Al_{0.3}Ga_{0.7}N$ EBL and a $B_{0.14}Al_{0.86}N$ EBL with Mg doping concentrations from $1 \times 10^{18}$, $1 \times 10^{19}$ to $1 \times 10^{20}$ /cm$^3$ at an injection current of 90 mA. The $E_{Fn}$ and $E_{Fp}$ are the quasi-Fermi energy levels of electrons and holes, respectively. As the Mg doping concentration in the EBL increases, the effective barrier height of the conduction band (defined as $\Phi_e = E_c - E_{Fn}$) can increase from 195 to 261 meV for $Al_{0.3}Ga_{0.7}N$ EBL structures in Fig. 2(a). While for $B_{0.14}Al_{0.86}N$ EBL structures, $\Phi_e$ can increase from 1.212 to 1.541 eV shown in Fig. 2(b). The significantly large $\Phi_e$ is because the lager conduction band offset between the $Al_{0.2}Ga_{0.4}N$ QB and $B_{0.14}Al_{0.86}N$ EBL. For both structures, the enhanced $\Phi_e$ suppresses the electron overflow out of the active region, indicating better capacities of confining electrons and reducing current leakage. As for the valence band, the effective barrier height (defined as $\Phi_h = E_{Fp} - E_v$) decreases with increasing Mg doping concentration, suggesting an enhanced hole injection capability for both structures. The modification of the barrier heights in the EBL region can be explained by the fact that the quasi-Fermi energy level of holes will become closer to the valence band edge as the Mg doping concentration increases. It is noted that the large $\Phi_e$ and diminutive $\Phi_h$ of $B_{0.14}Al_{0.86}N$ EBL are more favorable for the blocking of electrons and enhancing hole injection.

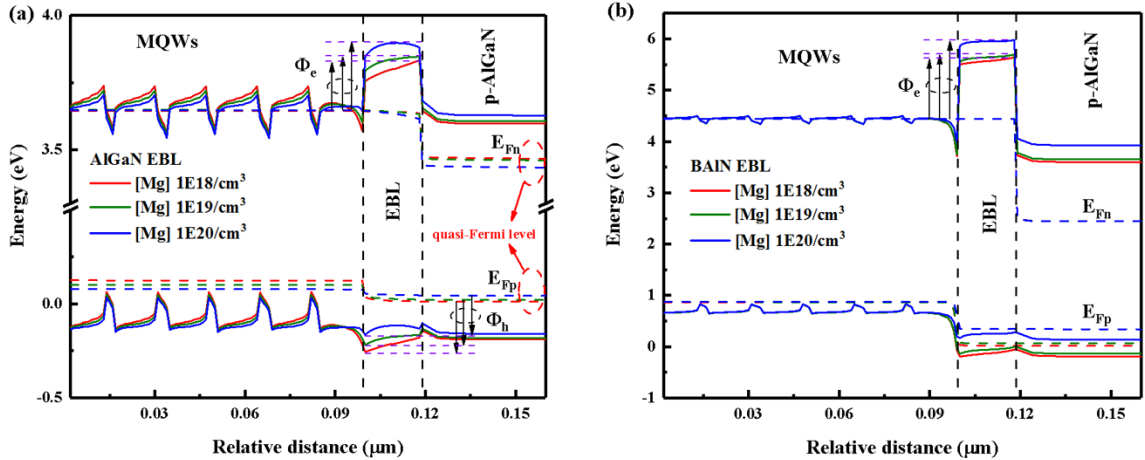

Fig. 2. Electronic band edge profiles at an injection current of 90 mA for the LED structures with (a) $Al_{0.3}Ga_{0.7}N$ EBL and (b) $B_{0.14}Al_{0.86}N$ EBL at various Mg doping concentrations.

To verify the analysis shown in Fig. Fig. 2. Electronic band edge profiles at an injection current of 90 mA for the LED structures with (a) $Al_{0.3}Ga_{0.7}N$ EBL and (b) $B_{0.14}Al_{0.86}N$ EBL at various Mg doping concentrations., we further study the electron and hole concentrations for both LED structures. The electron leakage in both structures decreases with increasing Mg doping concentrations, as shown in Fig. 3a. This phenomenon stems from the enlarged $\Phi_e$ as the increase of Mg doping concentration, which suppresses the electrons in the active region overflowing to the p-region. Comparing both LED structures, the LEDs with a $B_{0.14}Al_{0.86}N$ EBL show more significantly reduced electron leakage, even when the doping concentration reduces to a lower level such as $1 \times 10^{18}$ /cm$^3$ due to the relatively high $\Phi_e$. The electron and hole concentrations in the active region increase with the increase of p-doping levels for the LEDs with an Al$_{0.3}$Ga$_{0.7}$N EBL, resulting from the enlarged $\Phi_e$ and reduced $\Phi_h$, respectively (shown in Fig. 3b and 3c). Because of the large and small barrier heights of the conduction and valence bands, the carrier concentrations in QWs show less difference for the LEDs with a B$_{0.14}$Al$_{0.86}$N EBL. Based on the aforementioned results, a higher p-doping level in the Al$_{0.3}$Ga$_{0.7}$N or B$_{0.14}$Al$_{0.86}$N EBL is preferable for the enhancement of hole injection and blocking electrons. However, the B$_{0.14}$Al$_{0.86}$N EBL structures with low p-doping level can still possess high performance while the performance of Al$_{0.3}$Ga$_{0.7}$N EBL structures with low p-doping level is seriously deteriorated.

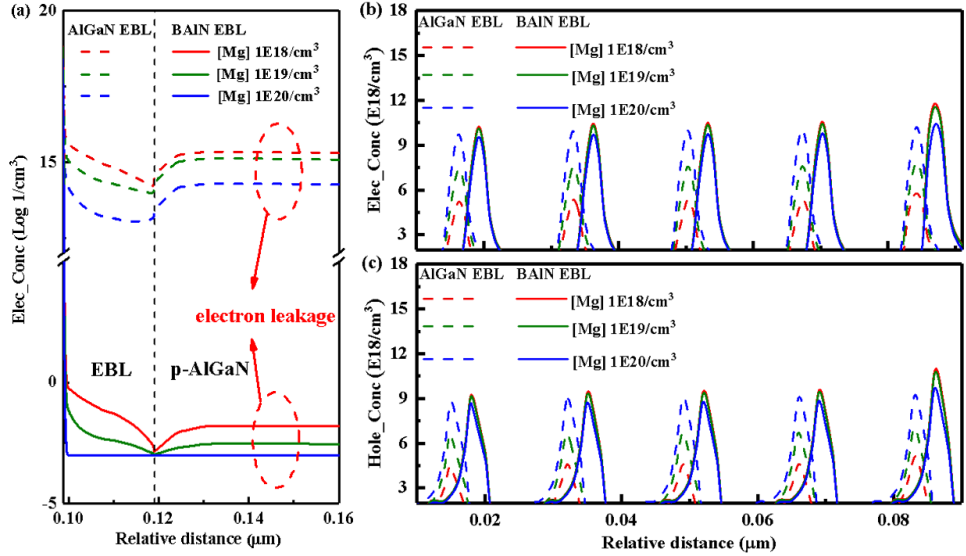

Fig. 3. (a) Electron leakage in the p-Al$_{0.2}$Ga$_{0.8}$N layer, (b) electron concentration, and (c) hole concentration in QWs at an injection current of 90 mA for the LED structures with Al$_{0.3}$Ga$_{0.7}$N and B$_{0.14}$Al$_{0.86}$N EBLs at various Mg doping levels. For better observation, we shift electron and hole concentration of B$_{0.14}$Al$_{0.86}$N EBL structures in (b) and (c) to the right by 3 nm.

Fig.Fig. represents the effect of the EBL with various doping levels on the IQE for both LED structures. The LEDs with an Al$_{0.3}$Ga$_{0.7}$N EBL exhibit an overall improvement in efficiency with increasing Mg doping concentration in the EBL. The increased IQE can be attributed to (1) the more holes activated with higher Mg doping concentration that can recombine with electrons in the active region, (2) the E$_{Fp}$ moves to the valence band, leading to a decreased $\Phi_h$ [40]. Moreover, the efficiency droop ratio is reduced to 8% with the highest p-doping level for the LEDs with an Al$_{0.3}$Ga$_{0.7}$N EBL, thanks to the enhanced hole injection and reduced electron leakage. The value of efficiency droop ratio is calculated using Equation (2), where the IQE$_{max}$ is the peak efficiency value and the IQE$_{90}$ is the value of efficiency at 90 mA. For the LEDs with a B$_{0.14}$Al$_{0.86}$N EBL, the IQE shows a slight increase with higher Mg doping concentration, ascribed to the superior electron blocking capability and nearly consistent hole injection capability for all B$_{0.14}$Al$_{0.86}$N EBLs with different doping levels. The efficiency droop ratio of the LEDs with a B$_{0.14}$Al$_{0.86}$N EBL can still sustain at 5%, even when the doping concentration reduces to 1×10$^{18}$ /cm$^3$, at which doping level the efficiency droop is significant for the LED with an Al$_{0.3}$Ga$_{0.7}$N EBL. All of the peak efficiency value for the LEDs with B$_{0.14}$Al$_{0.86}$N EBLs can reach as high as 74%, higher than that for any one of the LEDs with an Al$_{0.3}$Ga$_{0.7}$N EBL. Apparently, the efficiency of the LEDs with B$_{0.14}$Al$_{0.86}$N EBLs is less sensitive to the doping concentration, while high Mg doping is imperative for AlGaN EBL to acquire the high-efficiency UV LED.

$$\text{Efficiency droop ratio} = \frac{\text{IQE}_{max} - \text{IQE}_{90}}{\text{IQE}_{max}} \times 100\% \tag{2}$$

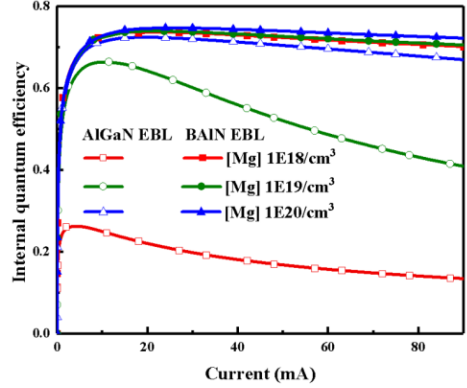

Fig. 4. Effect of p-doping level on IQE for Al$_{0.3}$Ga$_{0.7}$N and B$_{0.14}$Al$_{0.86}$N EBL LED structures at various Mg doping levels.

Fig. Fig. shows the comparisons of current-voltage (I-V) characterization curves and output powers. The forward voltage decreases with increasing Mg doping for both LED structures, resulting from the increased carrier concentration. However, the degree of change is different. The forward voltage of the LEDs with a B$_{0.14}$Al$_{0.86}$N EBL decreases substantially, whereas that of the LEDs with an Al$_{0.3}$Ga$_{0.7}$N EBL decreases slightly. The higher forward voltage of the LEDs with a B$_{0.14}$Al$_{0.86}$N EBL is due to that the flow of charge carrier is hindered by the higher barrier height. The LED with an Al$_{0.3}$Ga$_{0.7}$N EBL having the highest p-doping level shows a remarkable improvement when compared with that having the lowest doping level. Even with 10 times higher Mg doping concentration than the lowest-doping-level Al$_{0.3}$Ga$_{0.7}$N EBL, the output power of Al$_{0.3}$Ga$_{0.7}$N EBL LEDs can still increase by 208%. As for the output power of the LEDs with a B$_{0.14}$Al$_{0.86}$N EBL having the highest p-doping level, a maximum value of 235 mW can be achieved. Nearly the same output powers for the LEDs with a B$_{0.14}$Al$_{0.86}$N EBL are attributed to the perfect electron blocking capability and slightly increased hole injection. As expected, the LEDs with a B$_{0.14}$Al$_{0.86}$N EBL having relatively low doping still show enlarged output power compared to the LEDs with an Al$_{0.3}$Ga$_{0.7}$N EBL having the highest doping level. It further confirms that the low-doping-level B$_{0.14}$Al$_{0.86}$N EBL is still vitally significant in designing high-performance UV LEDs. Table 1 presents the wall-plug efficiency (WPE) of both LED structures. Although the LED with a B$_{0.14}$Al$_{0.86}$N EBL having the highest doping level shows a slight reduction in WPE as compared with the LEDs with an equally doped Al$_{0.3}$Ga$_{0.7}$N EBL, the LEDs with a B$_{0.14}$Al$_{0.86}$N having the lowest doping can still maintain the WPE at 51.2%.

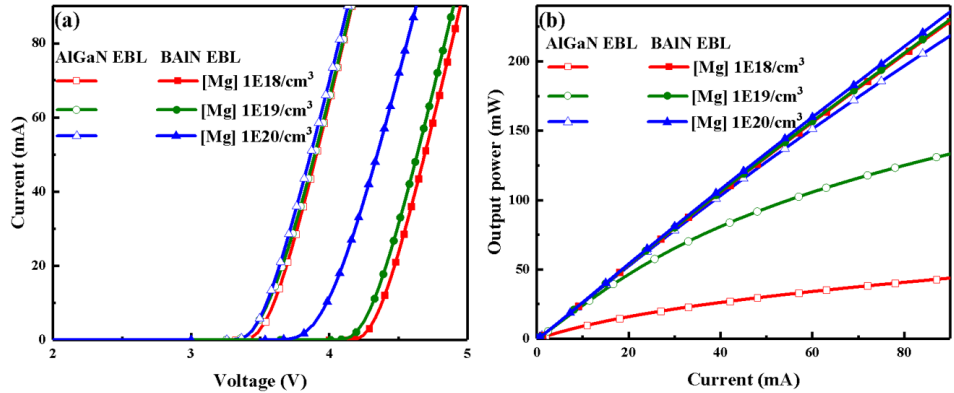

Fig. 5 (a) I-V characterization curve of Al$_{0.3}$Ga$_{0.7}$N and B$_{0.14}$Al$_{0.86}$N EBL LEDs (b) Effect of p-doping level on output power for Al$_{0.3}$Ga$_{0.7}$N and B$_{0.14}$Al$_{0.86}$N EBL LEDs.

**Table 1. WPE of Al$_{0.3}$Ga$_{0.7}$N and B$_{0.14}$Al$_{0.86}$N EBL structures with various doping concentrations at 90 mA.**

| Doping concentration of EBL | WPE of B$_{0.14}$Al$_{0.86}$N EBL structures | WPE of Al$_{0.3}$Ga$_{0.7}$N EBL structures |
|---|---|---|
| 1E18 /cm$^3$ | 51.2% | 11.6% |
| 1E19 /cm$^3$ | 52.2% | 35.5% |
| 1E20 /cm$^3$ | 56.7% | 58.6% |

We conclude that the performance of the LEDs with an $Al_{0.3}Ga_{0.7}N$ EBL shows heavy dependence on the p-doping level of EBL. As the doping concentration increases, the enhanced $\Phi_e$ holds back the transition of electrons to the p-region. Meanwhile, the reduced $\Phi_h$ promotes the hole injection to the active region. By comparing IQE and output power features for the LEDs with an $Al_{0.3}Ga_{0.7}N$ EBL, we propose that the p-doping level of $Al_{0.3}Ga_{0.7}N$ EBL is preferable to be improved for high-performance UV LEDs. As for the LEDs with a $B_{0.14}Al_{0.86}N$ EBL, we show that the performance is less dependent on the p-doping level of EBL. With the decrease of doping concentration in $B_{0.14}Al_{0.86}N$ EBL, the $\Phi_e$ shows a downward trend but maintains at a high level and $\Phi_h$ shows an upward trend but maintains at a low level, respectively. Meanwhile, the electron and hole concentrations, IQE, as well as output power show less difference in different doping levels in the EBL. We propose that the low-doping-level $B_{0.14}Al_{0.86}N$ EBL can still make a difference for acquiring high-performance UV LEDs.

## 4. Undoped EBL LED

It is well known that the generally adopted AlGaN EBL in UV LEDs will deteriorate the hole injection and introduce nonradiative recombination centers in MQWs [41]. Motivated by the diminutive valence band edge and large conduction band edge, we design the $B_{0.14}Al_{0.86}N$ EBL structures to avoid the p-doping issue. As discussed in part 3, the high p-doping level is not pre-requisite in designing EBL for LEDs after introducing the $B_{0.14}Al_{0.86}N$ EBL. To thoroughly demonstrate the potential of the undoped $B_{0.14}Al_{0.86}N$ EBL, we gather the effective barrier height of valence and conduction bands as a function of Mg doping concentration, as shown in Fig. . With low Mg doping concentration for both EBLs, the $\Phi_h$ and $\Phi_e$ barely decrease and increase with increasing doping concentration, respectively. Due to the relatively low activation energy of $B_{0.14}Al_{0.86}N$, the variations of $\Phi_e$ are more remarkable at the high p-doping level than that of $Al_{0.3}Ga_{0.7}N$. In contrast, the high doping level $B_{0.14}Al_{0.86}N$ EBL is expected to provide a larger $\Phi_e$ and a smaller $\Phi_h$. It is noteworthy that when the doping concentration reduces to zero, the $\Phi_e$ of $B_{0.14}Al_{0.86}N$ EBL is 6.20 times that of $Al_{0.3}Ga_{0.7}N$ EBL, while the $\Phi_h$ of $B_{0.14}Al_{0.86}N$ EBL is 0.76 times that of $Al_{0.3}Ga_{0.7}N$ EBL. Thus, the low doping level $B_{0.14}Al_{0.86}N$ EBL still has great potential in reducing electron leakage and increasing hole injection, which is meaningful for the proposing of undoped $B_{0.14}Al_{0.86}N$ EBL.

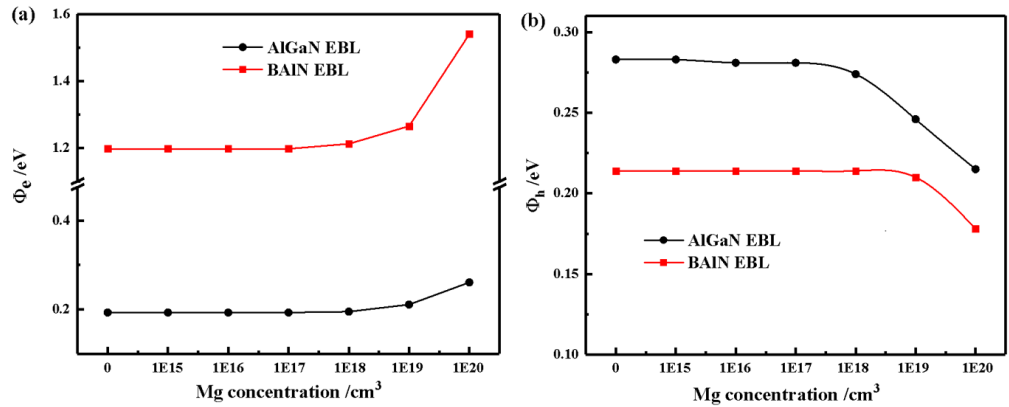

Fig. 6. Effective barrier height of (a) conduction and (b) valence band for variously doped $Al_{0.3}Ga_{0.7}N$ and $B_{0.14}Al_{0.86}N$ EBL at 90 mA.

To prove the superiority of undoped $B_{0.14}Al_{0.86}N$ EBL, we choose the undoped $B_{0.14}Al_{0.86}N$ EBL structure to compare with the $Al_{0.3}Ga_{0.7}N$ EBL structure with a high Mg doping concentration of $1\times10^{20}$ /cm³, which is deemed to be over the doping limit in the experiment [42]. As Fig. Fig. 7illustrates, the undoped $B_{0.14}Al_{0.86}N$ EBL structure shows enhancements of the electron and hole concentrations in the QWs because it facilitates the blocking of electrons and hole transport into QWs simultaneously. Compared with the $Al_{0.3}Ga_{0.7}N$ EBL structure, the increased electron concentration in the QWs of the undoped $B_{0.14}Al_{0.86}N$ EBL structure is due to that the larger $\Phi_e$ can lead to a declined electron leakage. In the meantime, because the relatively low $\Phi_h$ promotes the hole injection, the hole concentration in QWs for undoped $B_{0.14}Al_{0.86}N$ EBL structure shows an enhancement compared with the $Al_{0.3}Ga_{0.7}N$ EBL structure. When it comes to the radiative recombination, an enhancement indicates that a higher intensity of emitting light can be achieved by the utilization of undoped $B_{0.14}Al_{0.86}N$ EBL.

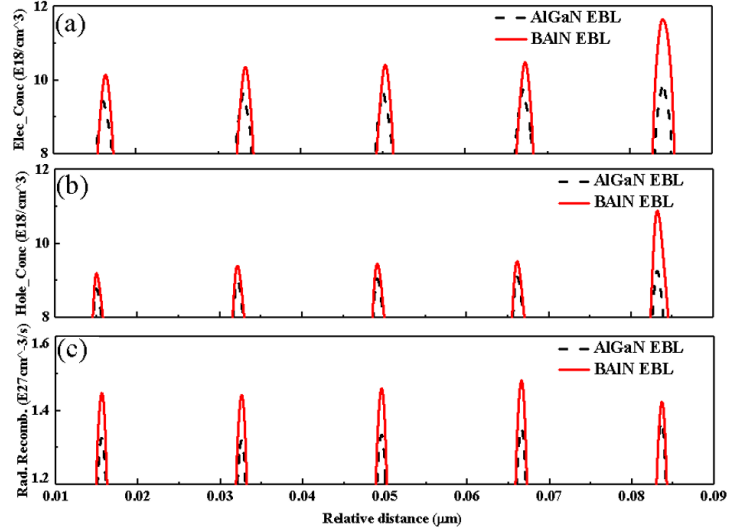

Fig. 7. (a) Electron concentration, (b) hole concentration, and (c) radiative recombination rate in QWs for $Al_{0.3}Ga_{0.7}N$ EBL structures and undoped $B_{0.14}Al_{0.86}N$ EBL structures at 90mA, respectively.

IQE is another vital parameter to evaluate the performance of the undoped LED. As shown in Fig. 8a, the undoped $B_{0.14}Al_{0.86}N$ EBL structure displays a slightly increased peak efficiency of 74% and reduces efficiency droop ratio at 90 mA compared with that of $Al_{0.3}Ga_{0.7}N$ EBL structure. Fig. 8b represents the output power characteristics for both structures. A slight improvement of output power is achieved by employing undoped $B_{0.14}Al_{0.86}N$ EBL to replace $Al_{0.3}Ga_{0.7}N$ EBL. Both of the improved IQE and enhanced output power are ascribed to the subdued electron leakage and enhanced hole injection for the undoped $B_{0.14}Al_{0.86}N$ EBL structure. The calculated WPE of the undoped $B_{0.14}Al_{0.86}N$ EBL structure is around 51%, which is slightly lower than the $Al_{0.3}Ga_{0.7}N$ EBL structure of 59%. The slightly low WPE is attributed to the large forward voltage induced by the higher band barrier height of the $B_{0.14}Al_{0.86}N$ EBL structure, and it will not lead to severe power dissipation.

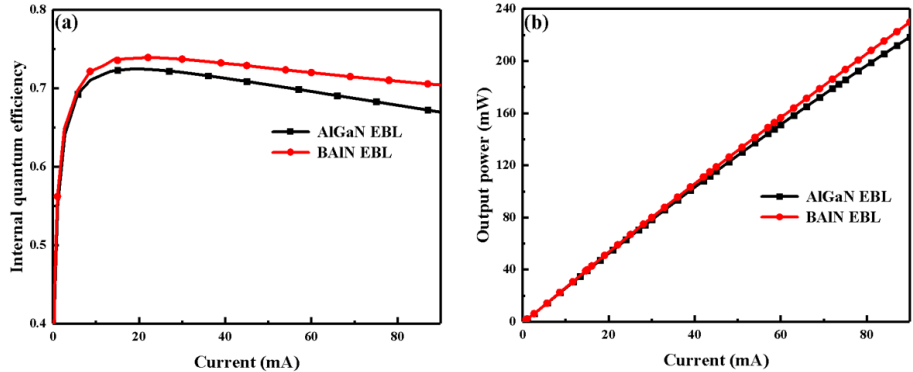

Fig. 8. (a) I-V characterization curve and (b) Output power features for $Al_{0.3}Ga_{0.7}N$ and undoped $B_{0.14}Al_{0.86}N$ EBL LEDs.

In summary, we propose an undoped $B_{0.14}Al_{0.86}N$ EBL structure to compare with the high-doping-level $Al_{0.3}Ga_{0.7}N$ EBL structure. The results show that the undoped $B_{0.14}Al_{0.86}N$ EBL structure still exhibits significant enhancements in blocking electrons and improving hole injection, because of the lager $\Phi_e$ and smaller $\Phi_h$. As for the characterization curve, the $B_{0.14}Al_{0.86}N$ EBL structure shows comparable IQE and mitigates efficiency droop as well as elevated output power density compared with the $Al_{0.3}Ga_{0.7}N$ EBL structure. Moreover, after the introducing of an undoped $B_{0.14}Al_{0.86}N$ EBL, the Mg diffusion issue also can be relieved.

## 5. Conclusion

The influence of various-doping-concentration $Al_{0.3}Ga_{0.7}N$ and $B_{0.14}Al_{0.86}N$ EBLs on the output features of UV LEDs has been systematically investigated. We reveal that the high doping level in $Al_{0.3}Ga_{0.7}N$ EBL is critical for the suppression of electron leakage and facilitates hole injection by elevated $\Phi_e$ and reduced $\Phi_h$. As a result, for LEDs with an $Al_{0.3}Ga_{0.7}N$ EBL with a doping concentration of $1\times10^{19}$ /cm$^3$, significant improvement in output power (208%) and enhanced IQE is achieved when compared with the LEDs with an $Al_{0.3}Ga_{0.7}N$ EBL at a doping concentration of $1\times10^{18}$ /cm$^3$. When adopting a $B_{0.14}Al_{0.86}N$ EBL instead of the $Al_{0.3}Ga_{0.7}N$ EBL, the performance of UV LEDs shows less deterioration with the decrease of doping concentration due to the intrinsic large conduction band offset and pimping valence band offset at $B_{0.14}Al_{0.86}N$/ $Al_{0.2}Ga_{0.8}N$ heterointerface. The comparison between the proposed undoped $B_{0.14}Al_{0.86}N$ EBL structure and the conventional highly-doped $Al_{0.3}Ga_{0.7}N$ EBL structure further demonstrates the potential of $B_{0.14}Al_{0.86}N$ EBL in improving the performance of UV LEDs. Based on these results, we propose an undoped $B_{0.14}Al_{0.86}N$ EBL structure, which is compatible with doping-free and high performance. By the employment of undoped $B_{0.14}Al_{0.86}N$ EBL, the p-doping issue in the conventional $Al_{0.3}Ga_{0.7}N$ EBL can be alleviated and therefore the epitaxy progress can be simplified. This work offered a novel sight to design high-performance UV LEDs without considering the high p-doping issue.

## Funding

KAUST Baseline BAS/1/1664-01-01, KAUST Competitive Research Grant URF/1/3437-01-01 and URF/1/3771-01-01 and OSR-2017-CRG6-3437.02, GCC REP/1/3189-01-01, National Key R&D Program of China 2016YFB0400800, National Natural Sciences Foundation of China 61875187, 61527814, 61674147, and U1505253, Beijing Nova Program Z181100006218007, and Youth Innovation Promotion Association CAS 2017157.